\documentclass[11pt,a4paper]{article}
\usepackage{jheppub}
\usepackage{mathrsfs}
\usepackage{amsmath}
\usepackage{mathrsfs}
\usepackage{color}
\usepackage{psfrag}
\usepackage{graphicx}
\usepackage{subfigure}

\newcommand{\bea}{\begin{eqnarray}}
\newcommand{\eea}{\end{eqnarray}}
\newcommand{\be}{\begin{equation}}
\newcommand{\ee}{\end{equation}}

\newcommand{\dsl}{\pa \kern-0.5em /}

\newcommand{\pa}{\partial}

\newcommand{\ba}{\begin{array}}
\newcommand{\ea}{\end{array}}
\newcommand{\bit}{\begin{itemize}}
\newcommand{\eit}{\end{itemize}}

\newcommand{\nd}{\textrm{d}}

\title{On matrix description of D-branes}

\author{Qiang Jia}

\affiliation{Interdisciplinary Center for Theoretical Study, 
 University of Science and Technology of China\\ Hefei, Anhui
 230026, China\\}

\emailAdd{skylo@mail.ustc.edu.cn}

\abstract{We study the low energy dynamics of a single Dp-brane carrying sufficient large number of D0-brane charges in type IIA theory. We assume the D-brane topology to be $R \times \mathcal{M}_{2n} $, where $\mathcal{M}_{2n}$ is a closed manifold admitting a symplectic structure.  We propose a new gauge fixing condition which eliminates the spatial gauge fluctuations on the Dp-brane. Using a conventional regularization method, one finds that the dynamics is characterized by D0-brane matrix description when the density of D0-branes is large enough. We also calculate the leading order interactions between two D2-branes carrying both electric and magnetic fluxes in matrix theory.}


\begin{document}

\begin{flushright}

USTC-ICTS-19-18\\

\end{flushright}

\maketitle
\flushbottom
\section{Introduction}

M-theory in the light-cone frame is conjectured to be characterized by matrix theory\cite{Banks:1996vh}, which is a super quantum mechanics with matrix degrees of freedom. Fundamental objects in M-theory can be described by matrix theory in terms of the degrees of freedom of matrices\cite{Banks:1996vh, Banks:1996nn, Castelino:1997rv}. Moreover, many examples show that the interactions between them calculated in matrix theory also coincide with those from 11-dimensional supergravity\cite{Aharony:1996bh,Lifschytz:1996rw,Lifschytz:1996bh,Chepelev:1997vx,Chepelev:1997fk}. Historically, matrix theory was first derived as an attempt to quantize supermembrane\cite{deWit:1988wri}. The theory is expected to have a discrete spectrum of states, which has a one-to-one correspondence with the elementary particle-like states (For example, graviton, 3-form field and gravitino in 11D supergravity) in spacetime. However, the spectrum of matrix theory is continuous and the interpretation of particle states is vague\cite{deWit:1988xki}. This puzzle was naturally resolved in matrix theory since it can be interpreted as a second quantized theory which captures the whole M-theory in light-cone frame.

We may also treat the matrix theory in type IIA superstring theory, in which the matrix degrees of freedom arise from the low energy dynamics of D0-branes\cite{Seiberg:1997ad, Sen:1997we}. In particular, various kinds of D-branes in type IIA superstring theory can be constructed in the matrix theory.  Motivated by the M2-brane quantization mentioned above, we study the similar process. We focus on a single Dp-brane with a topology $R \times \mathcal{M}_{2n} $, where $\mathcal{M}_{2n}$ is a closed manifold admitting a symplectic structure, and analyse the bosonic part of the dynamics. We turn on a time independent magnetic fluxes on the Dp-brane which give rise to $N$ D0-branes, and choose a special gauge to eliminate the spatial gauge fluctuations on the Dp-branes. We find that, after a regularization process, the dynamics of the D-brane is totally characterized by D0-branes when the density of D0-branes is large enough. This is not surprised since the Dp-brane looks more and more like a collection of D0-branes as the density grows.

We also study the leading interactions between a pair of D2-branes carrying both magnetic and electric fluxes on the worldvolume using matrix theory. The magnetic fluxes give rise to the charge of D0-branes, which is proportional to the dimension of matrix. The electric fluxes on D2-branes give rise to charge of F-strings which is related to an overall longitudinal velocities of the D0-branes bound states along the D2-brane direction in matrix theory. In order to make the matrix description valid, the density of D0-branes, or the strength of magnetic fluxes should be large enough. We find that matrix theory correctly reproduces the stringy results calculated in type IIA superstring theory truncated to the lightest open string modes. In particular, with a suitable choice of longitudinal velocity, there are open string pairs creating between the moving D0-branes, which is the analogy of open string pair production between D2-branes by electric fluxes.

This paper is organized as follows. In section 2, we propose a special gauge choice for the bosonic DBI action of a Dp-brane in type IIA theory, in which the remaining degrees of freedom will transform into those of D0-branes after regularization. In section 3, we briefly review the regularization process, which is the so-called Berezin-Toeplitz method\cite{Bordemann:1994,Ma:2008}, and reproduce the D0-brane action up to second order through regularization.  In section 4, we calculate the leading interaction between two parallel D2-branes carrying both electric and magnetic fluxes, using the matrix theory. We conclude in section 5.

\section{Gauge fixing of a Dp-brane}

In this paper, we mainly focus on the DBI action of a Dp-brane in type IIA theory,
\begin{equation}
	S_B = -T_p \int \nd^{p+1} \sigma \sqrt{-\det(G_{\alpha \beta}+2\pi \alpha' F_{\alpha \beta})},
\end{equation}
where $p \equiv 2n$ is even and $G_{\alpha \beta}$ is the pull back of spacetime metric:
\begin{equation}
	G_{\alpha \beta} = \eta_{\mu \nu} \frac{\partial X^{\mu}}{\partial \sigma^{\alpha}} \frac{\partial X^{\nu}}{\partial \sigma^{\beta}},\quad (\mu,\nu = 0,1,\cdots,9; \alpha,\beta=0,1,\cdots,2n),
\end{equation}
and $\sigma^{\alpha}$ are the coordinates of the worldvolume. Here we have set the spacetime flat and other spacetime background fields zero. We also assume the topology of Dp-brane to be $R \times \mathcal{M}_{2n}$, where $\mathcal{M}_{2n}$ is a closed symplectic manifold. To describe a Dp-brane carrying D0-brane charges, we turn on a time-independent background magnetic flux on the worldvolume:
\begin{equation}\label{backgroundF}
	F_{\alpha \beta}^B = \left(\begin{array}{cc}
	0&0\\
	0& 2\pi N\omega_{ab}
	\end{array}\right),
\end{equation}
where $N$ is related to the number of D0-brane charges on the worldvolume. We also require the 2-form $\omega$ be non-degenerate on $\mathcal{M}_{2n}$, that means the D0-brane charge density is nowhere vanishing on $\mathcal{M}_{2n}$, otherwise we do not expect the local dynamics of the Dp-brane can be characterized by D0-branes in the region devoid of D0-branes. Such requirement only applies for symplectic manifold, and we identify the 2-form $\omega_{ab}$ as the symplectic 2-form on $\mathcal{M}_{2n}$. We normalize the symplectic volume of $\mathcal{M}_{2n}$ to unity,
\begin{equation}
\int_{\mathcal{M}_{2n}} \frac{\omega_2^n}{n!} = 1,
\end{equation}
and the corresponding D0-brane charge can be read from the Chern-Simons term of the D-brane action as
\begin{equation}
 T_p\int_{\mathcal{M}_{2n}} \frac{1}{n!} [(2\pi \alpha') F^B]^n  = N^n T_0,
\end{equation}
where the tension of D0-brane is related to that of Dp-brane via $(2\pi)^{2n} \alpha'^n T_p = T_0$. Therefore the total number of D0-branes is $N^n$. In the following, we will work in the units $2\pi\alpha'=1$, and the tension relation is simply $(2\pi)^n T_p = T_0$. 
The full field strength $F_{\alpha \beta}$ is split into the background $F^{B}_{\alpha \beta}$ and the fluctuation $f_{\alpha \beta}$ as $F_{\alpha \beta} = F^{B}_{\alpha \beta} + f_{\alpha \beta}$.

There are two types of local symmetries of the worldvolume action. One is the diffeomorphism and the other is the U(1) gauge symmetry. Below we first focus on the infinitesimal coordinate transformation on an arbitrary coordinate patch on $\mathcal{M}_{2n}$, and the results between different patches can be glued together. Under the infinitesimal coordinate transformation, we have
\begin{equation}
	\sigma^{\alpha} \rightarrow \sigma^{\alpha} + \epsilon^{\alpha}(\sigma),\qquad (\alpha=0,1,2,\cdots,2n),
\end{equation}
where $\epsilon^{\alpha}$ is an infinitesimal vector field on $\mathcal{M}_{2n}$. The spacetime coordinates $X^{\mu}$s transform as worldvolume scalars:
\begin{equation}
	\delta X^{\mu} = -\epsilon^{\gamma} \partial_{\gamma}X^{\mu},
\end{equation}
while the field strength $F_{\alpha \beta}$ transforms as a worldvolume tensor:
\begin{equation}
	\delta F_{\alpha \beta} = -\partial_{\alpha}\epsilon^{\gamma} F_{\gamma \beta} - \partial_{\beta}\epsilon^{\gamma}F_{\alpha \gamma} - \epsilon^{\gamma} \partial_{\gamma} F_{\alpha \beta}.
\end{equation}
We wish to fix the background $F^B_{\alpha \beta}$ while doing a general coordinate transformation, which means we adsorb the variation of $ \delta F_{\alpha \beta}^B$ into $\delta f_{\alpha \beta}$ and define the transformation $\delta'$ as
\begin{equation}
	\delta' f_{\alpha \beta} \equiv \delta F_{\alpha \beta} =  \delta f_{\alpha \beta} + \delta F^B_{\alpha \beta}.
\end{equation}
Substituting the background \eqref{backgroundF} and rewriting the field strength in terms of gauge fluctuations $a_{\alpha}$ given by $f_{\alpha \beta} = \partial_{\alpha} a_{\beta} - \partial_{\beta} a_{\alpha}$, one can deduce the transformation of gauge fluctuations:
\begin{align} \label{transformation of a}
\delta' a_0 &= -\partial_0 \epsilon^{\alpha} a_{\alpha} - \epsilon^{\alpha} \partial_{\alpha} a_0, \nonumber \\
\delta' a_a &= -\partial_a \epsilon^{\alpha} a_{\alpha} - \epsilon^{\alpha} \partial_{\alpha} a_a - 2\pi N \epsilon^c \omega_{ca},
\end{align}
under the infinitesimal coordinate transformation. Here $a=1,2,\cdots,2n$, and the last term in the second line comes from the variation of background field.

The fluctuations $X^{\alpha}$s reflect the deformation of D-brane, which correspond to the excitations of open string propagating along the D-brane in the open string picture. If the energy is small compared to string scale, massive modes are frozen and we are left with the massless excitations, in which the longitudinal modes along the D-brane are usually designated as gauge degrees of freedom on the D-brane. Therefore, we have a double counting of some degrees of freedom if we include both the deformations $X^{\alpha}$ and the gauge field $A_{\alpha}$. Usually, one adopts a static gauge by setting $\sigma^{\alpha} = X^{\alpha}$ for a flat D-brane and keeps the gauge fields. Here we propose another gauge fixing condition in which we use the diffeomorphism of the worldvolume to gauge away the spatial gauge fluctuations $a_a$. Here we still choose
\begin{equation}
X^0 = \sigma^0 \equiv t,
\end{equation}
and for a fixed $t$, we are left with the spatial coordinate transformation of $\sigma^a$:
\begin{equation}
	\sigma^a \rightarrow \sigma^a + \epsilon^a(\sigma^a,t), \qquad (a=1,2,\cdots,2n),
\end{equation}
and we use them to gauge away the spatial fluctuations $a_a$ according to \eqref{transformation of a} such that\footnote{This can be easily carried out when $N$ is sufficiently large, where the variations are approximated as $\delta' a_a = -2\pi N \epsilon^c \omega_{ca}$ in the large $N$ limit. Therefore one may gauge away $a_a$ by simply setting $\epsilon^a = -\omega^{ac} a_c/(2\pi N) $. Further, unlike the background gauge potential,  there is no global obstruction for the gauge potential fluctuations. Therefore the result applies to the whole manifold $\mathcal{M}_{2n}$ by gluing different patches.}:
\begin{equation}
	a_a=0,\qquad (a=1,2,\cdots,2n).
\end{equation}

After utilizing the diffeomorphism, we are still left with a gauge transformation of $a_{\alpha}$ such that,
\begin{equation}
	\delta_{\Lambda} a_{\alpha} = -\partial_{\alpha} \Lambda(\sigma),
\end{equation}
which will break the gauge choice $a_a=0$. However, if we combine it with a spatial coordinate transformation
\begin{equation}
	\sigma^a \rightarrow \sigma^a - \frac{\omega^{ab}}{2\pi N} \partial_b \Lambda,
\end{equation}
which is a canonical transformation generated by $-\Lambda(\sigma)/(2\pi N)$, the combined transformation $\tilde{\delta}_{\Lambda} \equiv \delta_{\Lambda} + \delta'_{\Lambda}$ will leave $a_a$ zero. Here $\omega^{ab}$ is the inverse of the symplectic 2-form. Therefore we are left with a gauge symmetry $\tilde{\delta}_{\Lambda}$ which preserves the gauge choices and we will just omit the tilde and denote it as $\delta_{\Lambda}$.

In summary, we choose a gauge that,
\begin{equation}
	X^0 = \sigma^0\equiv t, \quad a_a=0\qquad (a=1,2,\cdots,2n),
\end{equation}
and a gauge transformation $\delta_{\Lambda}$ is left. Under the gauge transformation, one can verify that the remaining fields transform as
\begin{equation}
	\delta_{\Lambda} a_0 = \partial_0 \Lambda + \frac{1}{N} \left\{ \Lambda , a_0 \right\},
\end{equation}
and 
\begin{equation}
	\delta_{\Lambda} X^i = \frac{1}{N} \left\{\Lambda,X^i \right\},
\end{equation}
where $i=1,\cdots,9$ are the spatial indices of spacetime. Here the Poisson bracket is defined using the symplectic 2-form $2\pi \omega_2$ instead of $\omega_2$ as $\left\{ A,B \right\} \equiv \omega^{ab} \partial_a A \partial_b B / 2\pi$, which is convenient for later regularization. The partial gauge fixed action is then
\begin{equation} \label{action2}
S'_B = -T_p \int \nd \sigma^{p+1} \sqrt{-\det K_{\alpha \beta}},
\end{equation}
where $K_{\alpha \beta}$ consists three parts
\begin{align}
K_{\alpha \beta} =& \left(\begin{array}{cc}
-1 + \dot{X}^i \dot{X}^i&\dot{X}^i\partial_b X^i \\
\dot{X}^i\partial_a X^i&\partial_a X^i \partial_b X^i
\end{array}\right)\nonumber \\
&+\left(\begin{array}{cc}
0&\vec{0}\\ \vec{0}^T& 2\pi N \omega_{ab}
\end{array}\right) + \left(\begin{array}{cc}
0& -\partial_b a_0 \\ \partial_a a_0& 0_{2n\times 2n}
\end{array}\right).
\end{align}

Varying the action $S'_B$ will give the equations of motion. The equations of motion for $X^i$s are
\begin{equation}\label{EomX}
	\frac{\delta S'_B}{\delta X^i} = 0 \rightarrow \partial_{\alpha} \left(  \sqrt{-\det K}(K^{\alpha \beta} + K^{\beta \alpha}) \partial_{\beta}X_i \right) \equiv \Phi_i = 0,
\end{equation}
and the equation of motion for $a_0$ is
\begin{equation}\label{EomA}
	\frac{\delta S'_B}{\delta a_0} = 0 \rightarrow \partial_{\alpha} \left( \sqrt{-\det K} (K^{\alpha 0} - K^{0 \alpha}) \right)\equiv \Phi^0 = 0.
\end{equation}
Here $K^{\alpha \beta}$ is the inverse of $K_{\alpha \beta}$. Since we have chosen the gauge $X^0 = \sigma^0,a_a=0$ to eliminate $2n+1$ degrees of freedom, we have $2n+1$ additional equations associated to them:
\begin{equation}\label{Constrains}\left\{ \begin{array}{l}
\frac{\delta S_B}{\delta X^0} = 0 \rightarrow \partial_{\alpha} \left( \sqrt{-\det K}(K^{\alpha 0} + K^{0 \alpha}) \right)\equiv C_0 = 0\\
\frac{\delta S_B}{\delta a_a} = 0 \rightarrow \partial_{\alpha} \left( \sqrt{-\det K} (K^{\alpha a} - K^{a \alpha}) \right) \equiv C^a = 0
\end{array} \right. ,
\end{equation}
and they must be imposed as additional constraints. Actually, these $2n+1$ constrains are automatically satisfied providing the equations of motion in \eqref{EomX} and \eqref{EomA}, and one can verify the combinations:
\begin{align}
	C^a &= \frac{\omega^{ba}}{2\pi N}(\partial_b a_0 \Phi^0  - \partial_b X^i \Phi_i ), \nonumber \\  
	C_0 &= \partial_a a_0 C^a  + \partial_0 X^a \Phi_a.
\end{align}
Therefore we do not need to impose them as additional constraints. In fact, these constrains $\eqref{Constrains}$ are conservation equations of the energy momentum tensors on the worldvolume $\partial_{\alpha}T^{\alpha \beta} = 0$.

We rearrange $K_{\alpha \beta}$ as
\begin{align}
K_{\alpha \beta} &=\left(\begin{array}{cc}
-1 &\vec{0}\\ \vec{0}^T& 2\pi N \omega_{ab}
\end{array}\right) + \left(\begin{array}{cc}
\dot{X}^i \dot{X}^i& \dot{X}^i\partial_b X^i-\partial_b a_0 \\ \dot{X}^i\partial_a X^i+\partial_a a_0& \partial_a X^i \partial_b X^i
\end{array}\right).\nonumber \\
& = \left(\begin{array}{cc}
-1 &\vec{0}\\ \vec{0}^T& 2\pi N \omega_{ac}
\end{array}\right) \left( \mathbf{1} -  \left(\begin{array}{cc}
\dot{X}^i \dot{X}^i &\dot{X}^i\partial_b X^i-\partial_b a_0\\ \frac{\omega^{dc}}{2\pi N} (\dot{X}^i\partial_d X^i+\partial_d a_0)& \frac{\omega^{dc}}{2\pi N}(\partial_d X^i \partial_b X^i)
\end{array}\right)\right).
\end{align}
The discriminant is then evaluated as\footnote{Here we use the relation between the determinant and Pfaffian for any 2-form:
\begin{equation}
	\det A_{ab} = \textrm{pf}(A_2)^2, \nonumber
\end{equation}
where the Pfaffian of $A_2$ is defined via
\begin{equation}
	\frac{1}{n!} A_2^n = \textrm{pf}(A_{ab}) e_1 \wedge e_2 \wedge \cdots \wedge e_{2n}, \nonumber
\end{equation}
and $\{e_1,e_2,\cdots,e_{2n} \}$ is the standard basis on $\mathcal{M}_{2n}$.}:
\begin{equation}
	\sqrt{-\det K_{\alpha \beta}} = \textrm{pf}(2\pi N \omega_{ab}) \sqrt{\det \left[\mathbf{1} -  \left(\begin{array}{cc}
\dot{X}^i \dot{X}^i &\dot{X}^i\partial_b X^i-\partial_b a_0\\ \frac{\omega^{bc}}{2\pi N} (\dot{X}^i\partial_c X^i+\partial_c a_0)& \frac{\omega^{bc}}{2\pi N} (\partial_c X^i \partial_b X^i)\end{array}\right)\right]}.
\end{equation}

Usually, if the fluctuations in the determinant are small, one may expand the determinant order by order\footnote{The determinant is expanded using the formula:\begin{equation}
	[\det (\mathbf{1} + \Omega)]^{\frac{1}{2}} = 1+\frac{1}{2}\textrm{tr}\Omega-\frac{1}{4}\textrm{tr}\Omega^2 +\frac{1}{8} (\textrm{tr}\Omega)^2+\cdots. \nonumber
\end{equation}}, and the action is then:
\begin{align}
	S'_{B} =& - T_0 N^n \int \nd t \frac{\omega_2^n}{n!} \left( 1 - \frac{1}{2} (D_t X^i)^2 - \frac{1}{4N^2}\left\{X^i,X^j\right\}\left\{X^j,X^i\right\} - \frac{1}{8} (D_t X^i D_t X^i)^2 \right. \nonumber \\
	&  + \frac{1}{32N^4}\left(\left\{X^i,X^j\right\}\left\{X^j,X^i\right\} \right)^2 -\frac{1}{2N^2}D_t X^i \left\{X^i,X^j\right\}\left\{X^j,X^k\right\} D_t X^k  \nonumber \\
	&+\frac{1}{8N^2} D_t X^i D_t X^i \left\{X^i,X^j\right\}\left\{X^j,X^i\right\} -\frac{1}{8N^4} \left\{X^i,X^j\right\}\left\{X^j,X^k\right\}\left\{X^k,X^l\right\}\left\{X^l,X^i\right\}\nonumber \\
	& + \textrm{high orders}...
\end{align}
where the covariant derivative is defined by $D_t \equiv \partial_t - \left\{ a_0 , \cdot \right\}/N$, which transforms under gauge transformation as $D_t X^i \rightarrow \left \{ \Lambda , D_t X^i \right\}/N$. Note that although the gauge invariance of the original action \eqref{action2} is not manifest, it is easy to see that every terms in the expansion above are separately gauge invariant. To see that, the gauge variation on the RHS can be written as
\begin{equation}
\delta_{\Lambda} S'_B = -T_0 N^n \int \nd t \frac{\omega_2^n}{n!} \mathcal{L}_{v_{\Lambda}} \mathscr{L}[X^i,a_0],
\end{equation}
where $\mathcal{L}_{v_{\Lambda}}$ is the Lie derivative on $\mathcal{M}_{2n}$ generated by the Hamiltonian vector field $v^a =-\omega^{ab} \partial_b \Lambda/(2\pi N)$. Since the symplectic 2-form is invariant under the Lie derivative, we have
\begin{equation}
\delta_{\Lambda} S'_B = -T_0 N^n \int \nd t \mathcal{L}_{v_{\Lambda}} \left(\frac{\omega_2^n}{n!}  \mathscr{L}[X^i,a_0]\right),
\end{equation}
which vanishes as the RHS can be written into a total derivative on $\mathcal{M}_{2n}$\footnote{The Lie derivative of a p-form $\omega$ with respect to a vector field $X$ can be written as $\mathcal{L}_X \omega = i_X \nd \omega + \nd (i_X \omega)$, where $i_X \omega$ is the contraction of $\omega$ with $X$. In this case, $\omega$ is an $2n$-form on $\mathcal{M}_{2n}$ and $\nd \omega = 0$, therefore the Lie derivative is a total derivative. Moreover, since $i_X \omega$ is globally defined on $\mathcal{M}_{2n}$, the integral is zero.}.

\section{Dp-brane regularization}
In this section we adopt the regularization procedure studied in \cite{Bordemann:1994,Ma:2008} for a general closed symplectic manifold based on the Berezin-Toeplitz method, and show that the dynamics of Dp-brane can be reformulated into those of D0-branes when the density of D0-branes is large enough.

Considering a $2n$-dimensional closed symplectic manifold $(\mathcal{M}_{2n},2\pi \omega_2)$ with $2\pi \omega_2$ the symplectic 2-form. The Poisson bracket between any two functions on the manifold is defined by:
\begin{equation}
	\{ f , g \} = \frac{\omega^{ab}}{2\pi} \partial_a f \partial_b g,
\end{equation}
where $\omega^{ab}$ is the inverse of $\omega_{ab}$. The results of the Berezin-Toeplitz method is that: for any smooth function on the manifold $\mathcal{M}_{2n}$, we may associate a sequence of $N^n \times N^n$ matrices $T_{N}(f)$, such that as $N\rightarrow \infty$, the following properties hold:
\begin{align}
	&\lim_{N\rightarrow \infty} \parallel T_N(f) T_N(g) - T_N(fg) \parallel = 0, \nonumber \\
	&\lim_{N\rightarrow \infty} \parallel [T_N(f),T_N(g)] - \frac{i}{N} T_N(\left\{ f,g \right\})\parallel = 0,
\end{align}
where $\parallel \cdot \parallel$ is an arbitrary matrix norm. Further, it is proved in \cite{Ma:2014} that the symplectic integral can be replaced by matrix trace such that for any smooth functions $f_1,\cdots,f_m$ on $\mathcal{M}_{2n}$, we have
\begin{equation}
	\frac{1}{N^n} \textrm{Tr} (T_N(f_1) \cdots T_N(f_m)) = \int_{\mathcal{M}_{2n}} \frac{\omega^n}{n!} f_1 \cdots f_n + \mathcal{O}\left(\frac{1}{N}\right).
\end{equation}

Denote $\mathbf{X}^i \equiv T_N(X^i)$ and $\mathbf{A}_0 \equiv T_N(a_0)$ for sufficiently large $N$, one may do the following replacement according to the above discussion:
\begin{itemize}
\item $X^i(\tau,\sigma) \rightarrow \mathbf{X}^i(\tau,\sigma),\quad a_0(\tau,\sigma) \rightarrow \mathbf{A}_0(\tau,\sigma)$.
\item $\left\{\cdot,\cdot\right\} \rightarrow -i N \left[\cdot,\cdot\right]$.
\item $ \int_{\mathcal{M}_{2n}} \frac{\omega^n}{n!} \rightarrow \frac{1}{N^n}\textrm{Tr}$.
\end{itemize}
After doing that, one finds the action of D2-brane can be written into matrix form:
\begin{align}
	S_{\textrm{DBI}} =& -T_0 \textrm{STr} \int \nd t \left( 1 - \frac{1}{2} (D_t \mathbf{X}^i)^2  + \frac{1}{4}\left[\mathbf{X}^i,\mathbf{X}^j\right]\left[\mathbf{X}^j,\mathbf{X}^i\right]\right. \nonumber \\
	&- \frac{1}{8} (D_t \mathbf{X}^i D_t \mathbf{X}^i)^2  + \frac{1}{32}\left(\left[\mathbf{X}^i,\mathbf{X}^j\right]\left[\mathbf{X}^j,\mathbf{X}^i\right] \right)^2 \nonumber \\
	&+\frac{1}{2}D_t \mathbf{X}^i \left[\mathbf{X}^i,\mathbf{X}^j\right]\left[\mathbf{X}^j,\mathbf{X}^k\right] D_t \mathbf{X}^k \nonumber \\
	&-\frac{1}{8} D_t \mathbf{X}^i D_t \mathbf{X}^i \left[\mathbf{X}^i,\mathbf{X}^j\right]\left[\mathbf{X}^j,\mathbf{X}^i\right]\nonumber \\
	&-\frac{1}{8} \left[\mathbf{X}^i,\mathbf{X}^j\right]\left[\mathbf{X}^j,\mathbf{X}^k\right]\left[\mathbf{X}^k,\mathbf{X}^l\right]\left[\mathbf{X}^l,\mathbf{X}^i\right] + \textrm{high orders}...
\end{align}
which is exactly the leading and next leading order bosonic DBI action for N D0-branes\cite{Myers:1999ps}. Here we use the totally symmetric trace instead of the ordinary trace and the covariant derivative is $ D_t \equiv \partial_t + i [ \mathbf{A}_0,\cdot] $.

\section{Interaction between D2-branes in matrix theory}
In this section, we study the interactions between a pair of parallel D2-branes carrying both electric and magnetic fluxes in terms of matrix theory. We also work in the unit $2\pi \alpha' = 1$ in the following.
\subsection{Basic setup}
We adopt the convention given in \cite{Taylor:2001vb} and the full Lagrangian for matrix theory is
\begin{equation}\label{original Lagrangian}
L = \frac{T_0}{2} \textrm{Tr} \left[ D_0 X^i D_0 X^i + \frac{1}{2}[X^i,X^j]^2 + \theta^T (i \dot{\theta} - \gamma_i [X^i,\theta]) \right],
\end{equation}
where $i,j=1,2,\cdots,9$ and all fields are $N\times N$ matrices. The covariant derivative is given by $D_0 X^i = \partial_t X^i -i[A,X^i]$, which is different from the convention in the last section. We also include the fermionic fields and keep to the leading order compared to the Lagrangian in the last section.

We consider the matrix interactions on classical backgrounds satisfying the equations of motion and expand each of the matrices around the background. We split the bosonic fields $X^i$ in terms of the background $B^i$ and spatial fluctuations $Y^i$ such that:
\begin{equation}
	X^i = B^i + Y^i,
\end{equation}
and we also assume the backgrounds of the gauge field $A$ and fermionic fields $\theta$ vanish. In the following, we work with the background field method as in \cite{Lifschytz:1996rw,Becker:1997wh,Douglas:1996yp}, by choosing a background field gauge:
\begin{equation}
	D_{\mu}^{\textrm{bg}}A^{\mu} = \partial_t A -i [B^i,X^i] = 0.
\end{equation}
Following the conventional Faddeev-Popov gauge fixing procedure, we include the ghosts $C,\bar{C}$ and add a term $-(D_{\mu}^{\textrm{bg}}A^{\mu})^2$ to the action. Rotate to the Euclidean time according to $t \rightarrow -i\tau$, $A \rightarrow -i A$ and $L_E = -L$, the Lagrangian describing the fluctuations on the background is
\begin{align} \label{lagrangian}
L_{E} =& \frac{T_0}{2} \textrm{Tr} \left[ \partial_{\tau}Y^i \partial_{\tau}Y^i + \partial_{\tau} A \partial_{\tau} A + \partial_{\tau} \bar{C} \partial_{\tau} C + \theta^T \dot{\theta} - 2 [B^i,B^j][Y^i,Y^j] + 4i \dot{B}^i [A,Y^i]\right. \nonumber \\
& \left. - [Y^i,B^j][Y^i,B^j] - [A,B^j][A,B^j] - [\bar{C},B^j][C,B^j]  + \theta^T \gamma_i [B^i,\theta] \right],
\end{align}
where we only keep the quadratic interactions for the calculation of one-loop effective action. 

In this paper, we mainly focus on the interaction between two separated objects (D2-branes), which means we choose the background as
\begin{equation}
	B = \left[\begin{array}{cc}
(\textrm{Object 1})_{N_1\times N_1} & 0\\
0&(\textrm{Object 2})_{N_2 \times N_2}	
\end{array}	 \right],
\end{equation}
with each block represents a classical solution in matrix theory satisfying the equations of motion. Moreover, we only consider the off-diagonal degrees of freedom in the fluctuation matrices, as they represent the interactions between the two objects:
\begin{equation}\label{A,Y and theta}
	A = \left(\begin{array}{cc}
0&A\\
A^{\dagger}&0\\	
\end{array}	 \right),\quad
	Y^i = \left(\begin{array}{cc}
0&A_i\\
A^{\dagger}_i&0\\	
\end{array}	 \right),\quad
\theta = \left(\begin{array}{cc}
0&\chi\\
\chi^{\dagger}&0\\	
\end{array}	 \right),
\end{equation}
and 
\begin{equation}\label{ghost}
	C = \left( \begin{array}{cc}
0&C_2\\
C_1&0\\
	\end{array} \right), \quad \bar{C} = \left( \begin{array}{cc}
0&\bar{C}_1\\
\bar{C}_2&0\\
	\end{array} \right).
\end{equation}
Substituting the background fields and off-diagonal fluctuations into the Lagrangian \eqref{lagrangian}, we can write the Lagrangian into several parts according to the fluctuation fields.

We consider the presence of two parallel D2-branes extended in $X^1,X^2$ directions and with a separation $r$ in the $X^9$ direction, also on which we designate a flux configurations:\begin{equation}\label{fluxes}
	F_1 = \left[
	\begin{array}{ccc}
	0&f'_1&f_1\\
	-f'_1&0&g_1\\
	-f_1&-g_1&0
	\end{array}\right], \quad F_2 = \left[
	\begin{array}{ccc}
	0&f'_2&f_2\\
	-f'_2&0&g_2\\
	-f_2&-g_2&0
	\end{array}\right].
\end{equation}
Such a background configuration can be constructed in matrix theory. We summary the result here and leave the details in the appendix. The corresponding configuration in matrix theory is given by
\begin{equation}
B^1 = \left(\begin{array}{cc}
Q_1 + v_1 t &\ \\
\ & Q_2 + v_2 t\\
\end{array} \right),\quad
B^2 = \left(\begin{array}{cc}
P_1 - v'_1 t &\ \\
\ & P_2 - v'_2 t\\
\end{array} \right),\quad
B^9 = \left(\begin{array}{cc}
0 &\ \\
\ & r\\
\end{array} \right).
\end{equation}
Here $Q_1,P_1,Q_2,P_2$ are matrices satisfying
\begin{equation}
	[Q_1,P_1] = -2\pi i c_1,\quad [Q_2,P_2] = -2\pi i c_2,
\end{equation}
with dimension $N_1 \times N_1$ for $Q_1,P_1$ and $N_2 \times N_2$ for $Q_2,P_2$. Each of these two pairs describes a collection of $N_a (a=1,2)$ D0-branes extended in $X^1,X^2$ directions with a length $2\pi \sqrt{c_a N_a}$.  The non-commutative property reflects the nature that each set of D0-branes are non-trivially bounded together to form a D2-brane, where the charge of each D2-brane is given by
\begin{equation}
	Q_a^{\textrm{D2}} = 2\pi T_0 N_a c_a,
\end{equation}
in matrix theory. Moreover, each D2-brane has an overall longitudinal velocities\footnote{After rotating to the Euclidean time $t \rightarrow - i \tau$, the velocity becomes imaginary $v^* = -v$ such that the background field $B^i$ is still Hermitian.} given by $v_a,v'_a$, they give rise to the fundamental string charges smearing on the D2-branes via
\begin{equation}
	Q^{\textrm{F}}_a =  -2\pi T_0 N_a c_a v_a.
\end{equation}
Further, the parameters in matrix description are related to the fluxes on D2-branes via
\begin{equation}\label{g and c}
	g_a = \frac{1}{2\pi c_a},
\end{equation}
for magnetic fluxes and
\begin{equation}\label{f and v}
	v_a = \frac{f_a}{\sqrt{1+g_a^2 - f_a^2 -f'^2_a}} \approx \frac{f_a}{g_a},\quad v'_a = \frac{f'_a}{\sqrt{1+g_a^2 - f_a^2 -f'^2_a}} \approx \frac{f'_a}{g_a},
\end{equation}
for electric fluxes. Here the density of D0-branes should be large enough in order for the matrix description to be valid. And since the D0-brane density is proportional to the magnetic flux on the D2-brane, the magnetic fluxes on both D2-branes should go to infinity. Finally, via translations and rotation on the $X^1,X^2$ plane, one may set the background matrix to a standard form:
\begin{equation}\label{Backgound B}
B^1 = \left(\begin{array}{cc}
Q_1 + v t &\ \\
\ & Q_2 \\
\end{array} \right),\quad
B^2 = \left(\begin{array}{cc}
P_1  &\ \\
\ & P_2\\
\end{array} \right),\quad
B^9 = \left(\begin{array}{cc}
0 &\ \\
\ & r\\
\end{array} \right),
\end{equation}
with $v = \sqrt{(v_1 - v_2)^2+(v'_1-v'_2)^2}$ the relative longitudinal velocity between two D2-branes.

We follow the method in \cite{Aharony:1996bh} by switching to another representation of $P$ and $Q$:
\begin{equation} \label{Q,P,x,y}
	\left\{ \begin{array}{l}
	Q_1 = 2\pi c_1 x,\quad P_1 = i \frac{\partial}{\partial x} \\
	Q_2 = -i \frac{\partial}{\partial y},\quad P_2 = 2\pi c_2 y
	\end{array} \right. ,
\end{equation}
which preserve the commutator $[Q_1,P_1] = -2\pi i c_1$ and $ [Q_2,P_2] = -2\pi i c_2$. One should also change the off-diagonal matrices into functions, and trace into integral:
\begin{equation}
	A_{mn}(\tau) \rightarrow A(\tau,x,y),\quad \textrm{tr} \rightarrow \int \nd x \nd y,
\end{equation}
where the indices of the first block correspond to the variable $x$ and the second block correspond to $y$.

As pointed out in \cite{Aharony:1996bh}, there are two subtleties. The first is that in matrix theory, we use finite but large $N$ to calculate the spectrum, and then take $N$ to infinity. Therefore one should also put a finite but large cut-off on $x$ and $y$. However, it is difficult to perform an exact computation with such a cut-off. Instead, we calculate the spectrum and the wave functions on the entire axis and then regulate to finite by taking wave functions that are supported in the finite interval. This yields the correct overall fluxes dependence but might introduce numerical factors. 

The second subtlety is that, although we know exactly the class of matrices that we are integrating out, we do not know what class of functions after we rewrite the matrices into functions. Here we will simply take the the usual $L^2$ functions on $x$ and $y$, and it seems the most natural way is to take the basis of functions to be the eigenfunctions of $H_a \equiv Q_a^2 + P_a^2$, which are symmetric between $Q$ and $P$ as they are in the same position. They are harmonic oscillators with frequencies $ 4 \pi c_a$.

\subsection{The bosonic fluctuations}
We first analyse the bosonic fluctuations. Substitute the matrices background \eqref{Backgound B} into the Lagrangian and keep the off-diagonal degrees of freedom according to \eqref{A,Y and theta} and \eqref{ghost}, we have
\begin{align}\label{LA}
	L_A = T_0 \textrm{tr} & \left[  -A^{\dagger} \partial_{\tau}^2 A -2 A Q_2 A^{\dagger} Q_1 + A Q_2^2 A^{\dagger} + A^{\dagger} Q_1^2 A  \right. \nonumber \\
	&\left. -2 A P_2 A^{\dagger} P_1  + A P_2^2 A^{\dagger} + A^{\dagger} P_1^2 A + r^2 A^{\dagger} A \right. \nonumber \\
	&\left. + 2i v \tau  A^{\dagger} A Q_2 - 2i v \tau A^{\dagger}Q_1 A - v^2 \tau^2 A^{\dagger} A \right],
\end{align}
for gauge fluctuation and
\begin{align}\label{LY}
	L_Y = T_0 \textrm{tr} & \left[  -A_i^{\dagger} \partial_{\tau}^2 A_i -2 A_i Q_2 A_i^{\dagger} Q_1 + A_i Q_2^2 A_i^{\dagger} + A_i^{\dagger} Q_1^2 A_i  \right. \nonumber \\
	& -2 A_i P_2 A_i^{\dagger} P_1  + A_i P_2^2 A_i^{\dagger} + A_i^{\dagger} P_1^2 A_i + r^2 A_i^{\dagger} A_i \nonumber \\
	&\left. + 2i v \tau  A_i^{\dagger} A_i Q_2 - 2i v \tau A_i^{\dagger}Q_1 A_i - v^2 \tau^2 A_i^{\dagger} A_i \right],
\end{align}
for spatial fluctuations. Moreover, there is a mass term between $A$ and $A^i$ which is
\begin{equation}\label{massterm}
	L_{\textrm{mass}} = T_0 \textrm{tr}\left[ 2 i A_2^{\dagger} 2\pi(c_2-c_1) A_1 - 2 i A_1^{\dagger} 2\pi(c_2-c_1) A_2 + 2v A_1^{\dagger} A - 2v A^{\dagger} A_1 \right].
\end{equation}
Similarly, the ghost parts are
\begin{align}
L_{C_1} = \frac{T_0}{2} \textrm{tr} & \left[ -\bar{C}_1 \partial_{\tau}^2 C_1 - 2\bar{C}_1 Q_2 C_1 Q_1 + \bar{C}_1 Q_2^2 C_1 + \bar{C}_1 C_1 Q_1^2  \right.\nonumber \\
& \left. - 2\bar{C}_1 P_2 C_1 P_1 + \bar{C}_1 P_2^2 C_1 + \bar{C}_1 C_1 P_1^2 + r^2 \bar{C}_1 C_1 \right. \nonumber \\
& \left. -2iv \tau \bar{C}_1 C_1 Q_1 + 2i v \tau \bar{C}_1 Q_2 C_2 - v^2 \tau^2 \bar{C}_1 C_1 \right] ,
\end{align}
and 
\begin{align}
L_{C_2} = \frac{T_0}{2} \textrm{tr} & \left[ -\bar{C}_2 \partial_{\tau}^2 C_2 - 2\bar{C}_2 Q_1 C_2 Q_2 + \bar{C}_2 Q_1^2 C_2 + \bar{C}_2 C_2 Q_2^2  \right.\nonumber \\
& \left. - 2\bar{C}_2 P_1 C_2 P_2 + \bar{C}_2 P_1^2 C_2 + \bar{C}_2 C_2 P_2^2 + r^2 \bar{C}_2 C_2 \right. \nonumber \\
& \left. +2iv \tau \bar{C}_2 C_2 Q_2 - 2iv \tau \bar{C}_2 Q_1 C_2 - v^2 \tau^2 \bar{C}_2 C_2 \right].
\end{align}

Note that these Lagrangians share the same structure, therefore we mainly focus on the gauge fluctuation $L_A$, since other bosonic degrees of freedom behave similarly. Rewriting the matrices into functions as discussed in the last subsection, one finds the Lagrangian of gauge fluctuation is simply
\begin{equation}
	L_A = T_0 \int \nd x \nd y A^*(\tau,x,y)\left(-\partial_{\tau}^2 + H + r^2\right) A(\tau,x,y),
\end{equation}
where 
\begin{equation}
H = (-i \frac{\partial}{\partial y} + 2\pi c_1 x - iv \tau)^2 + (-i \frac{\partial}{\partial x} + 2\pi c_2 y)^2,
\end{equation}
with the velocity $v = \sqrt{(v_1 - v_2)^2+(v'_1-v'_2)^2}$. Therefore we find a Hamiltonian describing a charged particle moving on a two-dimensional plane with a background vector potential:
\begin{equation}
A_x = 2\pi c_2 y,\quad A_y = 2\pi c_1 x - iv \tau.
\end{equation}
However, since the Hamiltonian and eigenfunctions involve $\tau$, it is difficult to diagonalize the operator $-\partial_{\tau}^2 - H(\tau,x,y)$. Here we attempt to use a coordinate transformation of $\tau$ and $x$ to get rid of the $\tau$ dependence in the Hamiltonian. To do that, we adopt another gauge for the vector potential such that:
\begin{equation}
A_x = 0,\quad A_y = 2\pi (c_1-c_2) x - i v \tau,
\end{equation}
and the Hamiltonian becomes
\begin{equation}
H = \left(-i \frac{\partial}{\partial y} + 2\pi (c_1-c_2) x - iv \tau \right)^2 + \left(-i \frac{\partial}{\partial x}\right)^2.
\end{equation}
First, we redefine that
\begin{equation}\left\{\begin{array}{l}
 x' = \frac{2\pi (c_1 - c_2)}{\omega}x -\frac{i v }{\omega} \tau, \\
 \tau' = \tau,\quad y'=y,
\end{array}\right.
\end{equation}
where $\omega = \sqrt{(2\pi)^2(c_1-c_2)^2 - v^2}$ and we have
\begin{equation}
-\partial_{\tau}^2 - H(\tau,x,y) =- \frac{\partial^2}{\partial \tau'^2} + \frac{2 i v}{\omega} \frac{\partial}{\partial \tau'} \frac{\partial}{\partial x'}+\left(-i \frac{\partial}{\partial y'} + \omega x'\right)^2 + \left(-i \frac{\partial}{\partial x'}\right)^2.
\end{equation}
Then following with another coordinate transformation
\begin{equation}\left\{\begin{array}{l}
\tau '' = \frac{\omega}{2\pi(c_1 - c_2)} \tau' + \frac{i v}{2\pi(c_1-c_2)} x',\\
x''=x',\quad y''=y',
\end{array}\right.
\end{equation}
which will change the operator to that
\begin{equation}
-\partial_{\tau}^2 - H(\tau,x,y) = - \frac{\partial^2}{\partial \tau''^2}+\left(-i \frac{\partial}{\partial y''} + \omega x''\right)^2 + \left(-i \frac{\partial}{\partial x''}\right)^2.
\end{equation}
Notice that the overall Jacobian is unity during the transformation, since the Jacobian in each step is unity. Therefore we obtain a time independent Hamiltonian:
\begin{equation}
	H''(x'',y'') \equiv \left(-i \frac{\partial}{\partial y''} + \omega x''\right)^2 + \left(-i \frac{\partial}{\partial x''}\right)^2,
\end{equation}
which describes the Landau levels with a frequency $E_n = 2(n+\frac{1}{2})\omega$ and with $n=0,1,2\cdots$. Each Landau level has a degeneracy
\begin{equation}
	\mathcal{N} =\omega L_{x''} L_{y''},
\end{equation} 
where $L_{x''}$ and $L_{y''}$ are the length of the coordinates $x''$ and $y''$.

The other fields are analysed in the same method. In summary, we have 6 complex degrees of freedoms with energy level $E_n = 2(n+\frac{1}{2})\omega$ and two complex degrees of freedoms with energy level $E_n \pm 2\omega$ each, where the energy shift $\pm 2 \omega$ is due to the mass term \eqref{massterm}.

\subsection{Fermionic fluctuations}
The fermionic part is 
\begin{align}\label{Fermionic L}
L_{\theta} = T_0 \textrm{tr} & \left[ \chi^{\dagger} \partial_{\tau} \chi - \chi^{\dagger} \gamma_1 \chi Q_2 +\chi^{\dagger} \gamma_1 Q_1 \chi - i v \tau \chi^{\dagger} \chi  - \chi^{\dagger} \gamma_1 \chi P_2 +\chi^{\dagger} \gamma_1 P_1 \chi + r \chi^{\dagger} \gamma_9 \chi \right] ,
\end{align}
where $\gamma_i$ are the gamma matrices for SO(9) and we have omit the spinor indices for simplicity. Translating into the field theory language it becomes
\begin{equation}
L_{\theta} = T_0 \int \nd x \nd y \ \chi^* \left( \partial_{\tau} + (-i\frac{\partial}{\partial y} + 2\pi c_1 x - i v \tau)\gamma_1 + (-i\frac{\partial}{\partial x} + 2\pi c_2 y)\gamma_2 + r \gamma_9)  \right)\chi,
\end{equation} 
and the squared mass matrix is then
\begin{equation}
	M^2_{f} = H + 2\pi i (c_1 - c_2) \gamma_1 \gamma_2 + i v \gamma_1,
\end{equation}
where $H$ is the Hamiltonian given before and the discussion is parallel to that in the bosonic case. The eigenvalues of the last two matrices $2\pi i (c_1 - c_2)\gamma_1 \gamma_2 + i v \gamma_1$ are evaluated to be $\pm \omega$ with $\omega = \sqrt{(2\pi (c_2 - c_1))^2 - v^2}$ given before, which contributes to a energy shift. Therefore the energy levels are $E_n \pm \omega$ and each of them has 4 complex degrees of freedom, since only half of the fermions are viewed as creation operators.

\subsection{One-loop effective action}

The effective action is given by the logarithm of the partition function, whose one-loop contribution is given by
\begin{align}
	Z_{\textrm{one-loop}} =  \prod_{n} &[\det \left( -\partial^2_{\tau} + E_n \right)]^{-6}[\det \left( -\partial^2_{\tau} +  E_n + 2\omega \right)]^{-1} [\det \left( -\partial^2_{\tau} + E_n - 2\omega \right)]^{-1}  \nonumber \\
	& [\det \left( -\partial^2_{\tau} + E_n + \omega \right)]^{4} [\det \left( -\partial^2_{\tau} +  E_n - \omega \right)]^{4} ,
\end{align}
where the first line is the contributions from bosonic fields and the second line is those from fermionic fields. Using the integral representation of the determinant, the one-loop effective potential can then be evaluated as
\begin{align}\label{effective action 2}
W_{\textrm{one-loop}} = \log Z_{\textrm{one-loop}} & \sim T'' \mathcal{N}'' \int \frac{\nd s}{s^{\frac{3}{2}}} e^{-r^2 s}\frac{ \left(12 +4\cosh 2\omega s -16 \cosh \omega s \right)}{\sinh \omega s} \nonumber \\
& \sim T L_{x} L_{y} \omega \int \frac{\nd s}{s^{\frac{3}{2}}} e^{-r^2 s}\frac{ \sinh^4  \frac{\omega}{2} s}{\sinh \omega s},
\end{align}
where $T'' \mathcal{N}'' = T'' L_{x''} L_{y''} \omega = T L_{x} L_{y} \omega$, since the Jacobian is unity during the transformation.

We now roughly estimate the overall flux dependence given by the factor $L_x L_y \omega$. As discussed before, we take the basis of functions to be the eigenfunctions of the harmonic oscillators $H_a = Q_a^2 +P_a^2$ with frequency $4\pi c_a$, and the potential for each oscillator is $V_x = \frac{1}{2} (4\pi c_1)^2 x^2$ and $V_y = \frac{1}{2} (4\pi c_2)^2 y^2$. We truncate the eigenfunctions by considering those supported in the finite interval $L_x$ and $L_y$, and they are estimated by the states whose energies are lower than the height of the potential $\frac{1}{4} (4\pi c_1)^2 (\frac{L_x}{2})^2$ and $\frac{1}{4} (4\pi c_1)^2 (\frac{L_y}{2})^2$, since the wavefunctions with higher energies can spread outside the interval $[-\frac{L_x}{2},\frac{L_x}{2}]$ and $[-\frac{L_y}{2},\frac{L_y}{2}]$ significantly. Therefore the total numbers of truncated states are estimated as the height of the edge of the potential well $(\sim c^2 L^2)$ divided by the interval of energy level $(\sim c)$, which are approximated to the dimensions of the matrices in matrix theory:
\begin{equation}
	N_1 \sim c_1 L_x^2, \quad N_2 \sim c_2 L_y^2,
\end{equation}
which means the lengths of $x$ and $y$ are approximated as
\begin{equation}\label{LxLy}
L_x \sim \sqrt{\frac{N_1}{c_1}},\quad L_y \sim \sqrt{\frac{N_2}{c_2}}.
\end{equation}
Using the relations \eqref{g and c} and \eqref{f and v} given above, we may rewrite the degeneracy $L_x L_y \omega$ as
\begin{equation}
	L_{x} L_{y} \omega \sim \sqrt{A_1 A_2} \sqrt{(g_1 - g_2)^2 - (f_1 g_2 - f_2 g_1)^2 - (f'_1 g_2 - f'_2 g_1)^2 },
\end{equation}
where $A_a = 4\pi^2 c_a N_a$ are the area of two D2-branes, and $\sqrt{A_1 A_2}$ is the square root average. We may simply assume the areas of two D2-brane are the same, and the one-loop effective action is then
\begin{equation}
W_{\textrm{one-loop}} \sim V_3 \sqrt{(g_1 - g_2)^2 - (f_1 g_2 - f_2 g_1)^2 - (f'_1 g_2 - f'_2 g_1)^2  } \int \frac{\nd s}{s^{\frac{3}{2}}} e^{-r^2 s}\frac{ \sinh^4  \frac{\omega}{2} s}{\sinh \omega s},
\end{equation}
with $\omega$ given above and $V_3$ the spacetime volume of D2-brane. This is the correct effective potential (up to a numerical factor and truncated to the lightest open string modes) for a pair of D2-branes with flux configurations \eqref{fluxes} in the large $g_1,g_2$ limit.

Moreover, the integrand possesses infinite simple poles when the related velocity of the two D2-branes is large enough such that $v^2 > 4\pi^2 (c_1 - c_2)^2$, or equivalently, under a suitable choice of electric fluxes such that $(g_1 - g_2)^2 - (f_1 g_2 - f_2 g_1)^2 - (f'_1 g_2 - f'_2 g_1)^2<0$. These infinite simple poles give rise to an imaginary part of the effective action which indicate, as mentioned in the introduction, the open string pair production due to the electric fluxes on the D2-branes.

\section{Conclusion}
In the first part of this paper, we study the internal dynamics of a single D-brane. We focus on the bosonic part of a single D-brane in type IIA theory, assuming the topology is $R \times \mathcal{M}_{2n}$ where $\mathcal{M}_{2n}$ is a closed manifold admitting a symplectic structure. We study the limit in which the D-brane carries an sufficiently large, nowhere vanishing D0-brane density, and the symplectic structure is naturally related to the field strength. We choose a partial gauge that we set the worldvolume time equal to the spacetime time, and eliminate the gauge fluctuations in the spatial directions of worldvolume. After doing that, we are left with a residual gauge symmetry which corresponds to the gauge symmetry of D0-branes after regularization.  The constraint equations are automatically satisfied providing the equations of motion of the gauge fixed action. Following the Berezin-Toeplitz regularization process, we then show that the action can be rewritten into the low energy dynamics of $N$ D0-branes in the large $N$ limit. Therefore the (bosnoic part) dynamics of a single Dp-brane ($p>2$) in type IIA theory is characterized by D0-branes when the density of D0-brane is large enough. This is true since from the matrix theory, all type IIA objects should be characterized via D0-branes in the large $N$ limit, so as their dynamics. One may also say something about the transverse 5-brane issue in matrix theory. The tension of NS5-brane is proportional to $g_s^{-2}$ while that of D-branes are proportional to $g_s^{-1}$. Therefore, if NS5-brane can also be described by D0-branes, it must be a non-perturbative effect. That was verified in some papers discussing transverse 5-branes\cite{Ganor:1996zk, Maldacena:2002rb, Asano:2017xiy}. 

We also study the interactions between a pair of parallel D2-branes carrying fluxes in matrix theory. The magnetic fluxes give rise to charges of D0-brane, which should be large such that the matrix description is valid. The electric fluxes give rise to charges of fundamental string, and they correspond to the longitudinal velocity of D0-branes in matrix theory. We evaluate the one-loop effective action and find the matrix calculation gives the correct result, up to some constant factor and truncated to the lightest open string modes. Moreover, if the electric fluxes on the Dp-branes are chosen suitably, the effective potential possesses an imaginary part, which indicates the open string pair production. Similar discussions can be easily extended to other kinds of Dp-branes in type IIA superstring theory carrying both magnetic and electric fluxes.

\section*{Acknowledgements}

The author would like to thank Jianxin Lu, Zihao Wu and Xiaoying Zhu for discussion. The author acknowledge support by grants from the NSF of China with Grant No: 11775212 and 11235010.

\appendix

\section{D2-branes in matrix theory}
In this appendix, we review the construction of infinite extended BPS D2-branes in matrix theory\cite{Banks:1996vh}. The charge of D2-branes in matrix theory is given by\cite{Banks:1996nn,Taylor:2001vb}
\begin{equation}\label{D2charge}
	Q^{ij} = i T_0 \textrm {Tr} [X^i,X^j],
\end{equation}
which means that in order to construct a D2-brane, one should designate a non-trivial commutation relations between the coordinates $X^i$. In this paper, we consider D2-branes extended in the $X^1,X^2$ directions and let $X^1$ and $X^2$ satisfy
\begin{equation}
	[X^1,X^2] = -2\pi i c.
\end{equation}
The dimension of the matrices is $N\times N$ and the length of the D2-brane is related to the constant $c$ as $2\pi \sqrt{c N}$.\footnote{Such matrices can be constructed using 't Hooft matrices, see \cite{Banks:1996vh} and \cite{Taylor:2001vb} for details.}. Since $N$ is the number of D0-brane, this configuration is actually a D2-brane bounded with $N$ D0-branes. 

As a check, we calculate and verify the energy and charge of this D2-brane in matrix theory. The energy is read directly from the Lagrangian \eqref{original Lagrangian} as:
\begin{equation}
	E = -\textrm{Tr}\frac{T_0}{4} [X^i,X^j]^2 = \frac{N T_0}{2} (2\pi c)^2.
\end{equation}
On the other hand, the energy of a D2-D0 bound state is:
\begin{equation}\label{D2D0 energy}
	E' = \sqrt{T_0^2 N^2 + M_{D2}^2} \approx N T_0 + \frac{M_{D2}^2}{2 N T_0} + \mathcal{O}\left(\frac{1}{N^3}\right),
\end{equation}
where the first term is the total mass of $N$ D0-branes, which is subtracted in the matrix theory Lagrangian \eqref{original Lagrangian}. The second one is the energy related to the D2-brane and we have the relation in the large $N$ limit:
\begin{equation}
\frac{M_{D2}^2}{2 N T_0} = \frac{N T_0}{2} (2\pi c)^2 \rightarrow M_{D2} = 2\pi T_0 N c = T_2 A_{\textrm{D2}}.
\end{equation}
Here the area $A_{\textrm{D2}}$ of the D2-brane is $(2\pi)^2 N c$, therefore the tension of the D2-brane is $T_0 / 2\pi$, which is correct in the unit $2\pi \alpha' = 1$. Further, the D2 charge is given by \eqref{D2charge} as
\begin{equation}
	Q^{12} = i T_0 \textrm{Tr} \textrm [X^1,X^2]= 2 \pi T_0 N c,
\end{equation}
and is equal to the mass of the D2-brane, which is also true as a consequence of its BPS property.

The D0-branes dissolve into the D2-brane as magnetic flux, and we designate the field strength as
\begin{equation}
	F_{12} = - F_{21} = g.
\end{equation}
One should also match the magnetic flux $g$ on the D2-brane with the parameter $c$ in the matrix configuration. The corresponding D0-brane charge is given by the Chern-Simons term of the D2-brane as
\begin{equation}
	T_2 \int F_2 = A_{\textrm{D2}} T_2 g = N T_0.
\end{equation}
Using $T_2 = T_0 / 2\pi$, one may find the relation between magentic flux $g$ and parameter $c$:
\begin{equation}
	g = \frac{1}{2\pi c}.
\end{equation}
In order for the matrix description to be valid, the D0 charge density should be large, which means we work in the limit $c\ll 1$, or equivalently, $g \gg 1$.

We will next consider the electric fluxes on the worldvolume of D2-brane and its correspondence in matrix theory. We turn on electric fluxes in addition to the magnetic flux on the worldvolume and the field strength reads
\begin{equation}\label{Fieldstrenght}
	F_{\alpha \beta} = \left[
	\begin{array}{ccc}
	0&f'&f\\
	-f'&0&g\\
	-f&-g&0
	\end{array}\right].
\end{equation}
The electric fluxes on a D2-brane will contribute to effective F-string currents which can be read from the D2-brane action as:
\begin{equation}
	j^{\mu \nu} = \frac{1}{T_F} \frac{\delta S_{\textrm{D2}}}{\delta B_{\mu \nu}},
\end{equation}
where $S_{\textrm{D2}}$ is the conventional DBI action for D2-brane, $B_{\mu \nu}$ is the 2-form which couples F-string and $T_F$ is the tension of F-string. Since we are considering infinite extended D2-brane in the $X^1,X^2$ direction, we choose the static gauge such that $\sigma^{\alpha} = X^{\alpha} (\alpha=0,1,2)$, where $\sigma^{\alpha}$ are the worldvolume coordinates. We are interested in the charge density $j^{0i}$ of F-string along the D2-brane and for the present field configurations they are\footnote{There should also be a $\delta^7(x-x_0)$ in the charge density, where $x_0$ is the position of the D2-brane in the transverse directions, since the densities are localized on the D2-brane. Here we have integrated the transverse directions out.}
\begin{equation}
j^{01} = -\frac{T_2 f'}{T_F \sqrt{1+g^2-f^2-f'^2}}, \quad j^{02} = -\frac{T_2 f}{T_F \sqrt{1+g^2 - f^2 -f'^2}}.
\end{equation}
On the other hand, the charge of F-string in matrix theory is given by \cite{Banks:1996nn,Taylor:2001vb},
\begin{equation}\label{Fcharge}
	Q^i = i T_0 \textrm{Tr}\left([X^i,X^j]\dot{X^j} + [[X^i,\theta^{\alpha}]\theta^{\alpha}] \right).
\end{equation}
Since we have set the fermionic backgrounds zero, the second term vanish. In order to produce a non-zero F-string charges along the $X^1,X^2$ directions, we need a configurations with non-zero commutator $[X^1,X^2]$ and velocities $\dot{X^1}$ and $\dot{X^2}$. Therefore we add overall longitudinal velocities to D0-branes along the $X^1,X^2$ directions as
\begin{equation}
	\dot{X^1} = v,\quad \dot{X^2} = -v',
\end{equation}
and the F-string charge from matrix theory is then
\begin{equation}
	Q^1 = -2\pi T_0 N c v',\quad Q^2 = -2 \pi T_0 N c v,
\end{equation}
and the charge densities are those divided by area of the D2-brane, which are
\begin{equation}
	q^1 = -\frac{T_0 v'}{2\pi},\quad q^2 = -\frac{T_0 v}{2\pi}.
\end{equation}
Equate $j^{0a}$ and $q^a$, one finds the relations between electric fluxes $f,f'$ and the D0-brane longitudinal velocities $v,v'$:
\begin{equation}
	v = \frac{f}{\sqrt{1+g^2 -f^2 - f'^2}},\quad v' = \frac{f'}{\sqrt{1+g^2 -f^2 - f'^2}},
\end{equation}
where we have also used the tension relations $T_0 = 2\pi T_2$ and $T_F = (2\pi \alpha')^{-1} = 1$ in the unit $2\pi \alpha'=1$. Since the magnetic flux goes to infinity, the velocities are approximated as 
\begin{equation} 
 v = \frac{f}{g},\quad v' = \frac{f'}{g}.
\end{equation}

Therefore, the D2-brane extended in the $X^1,X^2$ direction carrying fluxes as \eqref{Fieldstrenght} can be described in matrix theory as
\begin{equation}
	X^1 = Q + v t,\quad X^2 = P - v't,
\end{equation}
where $Q,P$ are $N\times N$ matrices satisfying $[Q,P] = -2\pi i c$, and the parameters $c,v,v'$ are related to the fluxes in \eqref{Fieldstrenght} as $g=1/(2\pi c), v = f/g$ and $v'=f'/g$. As a check, we will again compare the energy calculated in both side. The energy calculated in matrix theory is read directly from the Lagrangian \eqref{original Lagrangian} as:
\begin{equation}\label{energy in matrix theory}
	E = T_0 \textrm{Tr} \left[ \frac{1}{2} \dot{X}^i \dot{X}^i - \frac{1}{4}[X^i,X^j]^2 \right] = \frac{N T_0}{2} (2\pi c)^2 + \frac{NT_0}{2} (v^2 + v'^2).
\end{equation}
On the other hand, the Lagrangian for the D2-brane is
\begin{equation}
	\mathcal{L}_{D2} = -T_2 \sqrt{ - \det (G_{\alpha \beta} + F_{\alpha \beta})},
\end{equation}
with $G_{\alpha \beta}$ the induced metric and $F_{\alpha \beta}$ the field strength given in \eqref{Fieldstrenght}. Here we still work in static gauge such that $\sigma^{\alpha} = X^{\alpha} (\alpha=0,1,2)$. The Hamiltonian is then\footnote{Note that the $X^i$ below are different from the $X^i$ in the matrix theory above.}
\begin{equation}
	\mathcal{H}_{D2} = \dot{A}_a \frac{\partial \mathcal{L}_{\textrm{D2}}}{\partial (\dot{A}_a)} + \dot{X}^i \frac{\partial \mathcal{L}_{\textrm{D2}}}{\partial (\dot{X}^i)} - \mathcal{L}_{\textrm{D2}},
\end{equation}
where $a=1,2$ and $i=3,\cdots,9$. In the present case we have $\dot{X}^i =0, \dot{A}_1= f'$ and $\dot{A}_2 = f$. Therefore the 
energy density can be evaluated as
\begin{equation}
	\mathcal{H}_{\textrm{D2}} = T_2 \left[ \frac{f^2 + f'^2}{\sqrt{1+g^2-f^2-f'^2}} + \sqrt{1+g^2-f^2-f'^2} \right],
\end{equation}
which is approximated as
\begin{equation}
	\mathcal{H}_{\textrm{D2}} \approx T_2 \left( g +\frac{1}{2g} + \frac{f^2 + f'^2}{2g} \right) = \frac{T_0}{(2\pi)^2 c} + \frac{T_0 c}{2} + \frac{T_0(v^2+v'^2)}{8\pi^2 c},
\end{equation} 
when the magnetic flux $g$ is large enough. Here we have used the relations $g=1/(2\pi c), v = f/g, v'=f'/g$ and $2\pi T_2 = T_0$ in the RHS. The total energy is then given by
\begin{equation}
E'= A_{D_2} \mathcal{H}_{\textrm{D2}} = N T_0 + \frac{N T_0}{2} (2\pi c)^2 + \frac{NT_0}{2} (v^2 + v'^2),
\end{equation}
where $A_{\textrm{D2}} = (2\pi)^2 N c$ the area of the D2-brane. Note that the first term is the again the total mass of $N$ D0-branes, which is subtracted in the matrix theory, and the second and third terms are equal to those calculated in the matrix theory given above in \eqref{energy in matrix theory}.

\end{document}